\documentclass[floats,floatfix,showpacs,amssymb,prl,twocolumn,superscriptaddress,nofootinbib]{revtex4-1}

\newlength{\figw} 
\usepackage{notoccite}

\usepackage{graphicx,epsf, epsfig, amssymb}
\usepackage{bm}
\usepackage{longtable,tabularx}
\usepackage{color}
\usepackage[breaklinks]{hyperref}
\usepackage{amsfonts,amsmath,amssymb,mathrsfs,gensymb}
\usepackage{natbib}
\usepackage{array}
\usepackage{multirow}
\usepackage{rotating,array}
\usepackage{graphicx}
\usepackage{soul}
\usepackage[normalem]{ulem}
\usepackage[T1]{fontenc}

\definecolor{linkcolor}{rgb}{0.0,0.3,0.5}
\definecolor{venetianred}{rgb}{0.78, 0.03, 0.08}
\hypersetup{colorlinks=true,linkcolor=linkcolor,citecolor=linkcolor,filecolor=linkcolor,urlcolor=linkcolor}

\newcommand{\rmi}{\mathrm{i}} 
\newcommand{\Jmath}{\ensuremath{\hat{\jmath}}} 
\newcommand{\smallcaps}[1]{{\textsc{#1}\! }} 

\def\apj{\rm{ApJ}}
\def\apjl{\rm{ApJ}}

\def\aap{\rm{A\&A}}

\def\mnras{\rm{MNRAS}}

\def\prd{\rm{Phys.~Rev.~D}}

\def\nat{\rm{Nature}}

\begin{document}

\title{An astrometric search method for individually resolvable gravitational wave sources with Gaia}

\author{Christopher J. Moore}
\email{cjm96@cam.ac.uk}
\affiliation{Department of Applied Mathematics and Theoretical Physics, Centre for Mathematical Sciences, University of Cambridge, Wilberforce Road, Cambridge CB3 0WA, UK}

\author{Deyan Mihaylov}
\affiliation{Intitute of Astronomy, University of Cambridge, Madingley Road, Cambridge CB3 0HA, UK}

\author{Anthony Lasenby}
\affiliation{Astrophysics Group, Cavendish Laboratory, J J Thomson Avenue, Cambridge CB3 0HE, UK}
\affiliation{Kavli Institute for Cosmology, Madingley Road, Cambridge CB3 0HA, UK}

\author{Gerard Gilmore}
\affiliation{Intitute of Astronomy, University of Cambridge, Madingley Road, Cambridge CB3 0HA, UK}


\date{\today}

\begin{abstract}
Gravitational waves (GWs) cause the apparent position of distant stars to oscillate with a characteristic pattern on the sky. Astrometric measurements (e.g.\ those made by Gaia) therefore provide a new way to search for GWs. The main difficulty facing such a search is the large size of the data set; Gaia observes more than one billion stars. In this letter the problem of searching for GWs from individually resolvable supermassive black hole binaries using astrometry is addressed for the first time; it is demonstrated how the data set can be compressed by a factor of more than $10^6$, with a loss of sensitivity of less than $1\%$. This technique is successfully used to recover artificially injected GWs from mock Gaia data. Repeated injections are used to calculate the sensitivity of Gaia as a function of frequency, and Gaia's directional sensitivity variation, or \emph{antenna pattern}. Throughout the letter the complementarity of Gaia and pulsar timing searches for GWs is highlighted.
\end{abstract}

\maketitle 

\noindent{\bf \em Introduction~--~}
The first detection of gravitational waves (GWs) from merging stellar mass black holes in the frequency range $(10\textrm{--}10^{4})\textrm{Hz}$ has recently been achieved by Advanced LIGO \cite{PhysRevLett.116.061102}. Advanced LIGO can detect binaries with total mass up to $\gtrsim\!160M_{\odot}$ \cite{2016ApJ...818L..22A}; however, heavier supermassive black hole binaries radiate at lower frequencies and are inaccessible to ground-based instruments. Observing GWs from these massive systems would shed light on the black hole mass function and the coalescence process of the host galaxies and is therefore a target for both current and future searches. There is progress towards a space-based detector, called LISA, which will detect merging binary black holes in the mass range $(10^{5}\textrm{--}10^{7})M_{\odot}$ out to redshifts $z\!\lesssim\!20$ \cite{2013arXiv1305.5720C}.

Other ongoing efforts include pulsar timing arrays (PTAs) which utilise the precise timing of pulsars to detect GWs with ${10^{-9}\lesssim f/\textrm{Hz}\lesssim 10^{-7}}$. Such GWs may be generated in the early inspiral of a binary in the mass range $(10^{7}\textrm{--}10^{10})M_{\odot}$. A GW passing over the Earth--pulsar system induces a Doppler shift to the pulsar which in turn affects the pulse arrival times at the Earth. By making a number of time-of-arrival measurements over a timespan $T$ (individual measurements separated by $\delta t$) PTAs achieve sensitivity to GWs in the range ${1/T \! \lesssim \!f\! \lesssim \! 1/2\delta t}$ \cite{2015CQGra..32e5004M}. Current PTAs include {\smallcaps{NanoG}\!rav} \cite{2013CQGra..30v4008M}, \smallcaps{Epta} \cite{2013CQGra..30v4009K}, \smallcaps{Ppta} \cite{2013CQGra..30v4007H}, and the combined \smallcaps{Ipta} \cite{2013CQGra..30v4010M}.

It is also possible to detect GWs using astrometry. The passage of a GW over the Earth--star system induces a deflection to the apparent position of a the star which depends on the components of the metric perturbation projected along the line-of-sight. By making repeated astrometric measurements of many objects across the sky and recording their changing position it is possible to identify the characteristic deflection pattern of a GW and turn an astrometric data set into a nHz GW observatory.

The ESA space-astrometry mission Gaia \cite{refId0Gaia}, in operation since 2014, is providing an all-sky astrometric and photometric map of over $10^{9}$ stars. Gaia will operate for $5\textrm{--}10$ years, making around 80 observations (in $5\,\textrm{years}$) per source, delivering proper motion accuracy of $20\,\mu\textrm{as}\,\textrm{yr}^{-1}$ at magnitude 15, degrading to $300\,\mu\textrm{as}\,\textrm{yr}^{-1}$ at the magnitude limit of 20.7.

The sensitivity bandwidth of Gaia is set by the measurement timings (similar to PTAs); Gaia is sensitive GWs with ${f\!\gtrsim\! 1/T}$. Gaia and PTAs can search for GWs from several sources, including monochromatic GWs from resolvable circular binaries, a stochastic background from the superposition of many binaries (or from cosmic string networks \cite{1981PhRvD..24.2082V} or early universe perturbations \cite{1976PZETF..23..326G}), or bursts with memory \cite{2015MNRAS.446.1657W,2015ApJ...810..150A}. The astrometric analysis of a nearly monochromatic wave is considered here; for example, GWs from a supermassive black hole binary in the early post--Newtonian inspiral stage of its evolution.

This letter begins by describing the astrometric effect of a GW. The data analysis principles that have been developed to search for monochromatic GWs with Gaia are then summarised and it is demonstrated how the data may be greatly compressed with little loss in sensitivity. A number of mock Gaia data sets are used to demonstrate the reliable recovery of GWs, quantify the accuracy with which the wave parameters (amplitude, frequency, etc.) can be measured, and quantify the sensitivity of Gaia both as a function of frequency and sky position.

\medskip
\noindent{\bf \em The astrometric response to GWs~--~}
Astrometric measurements of any distant objects may be used to detect GWs; for simplicity the term ``star'' is used to refer to all such objects. The telescope used for the astrometric measurements is not at rest (Gaia is orbiting about the L2 point) and it will be necessary to correct for the telescope's motion; it will be assumed that the necessary corrections have been made and the term ``Earth'' is used to refer to an idealised stationary observer.

The possibility of detecting GWs via astrometric deflections was first suggested in \cite{1990NCimB.105.1141B}. The astrometric deflection of a distant star was first derived in \cite{1996ApJ...465..566P} (also see \cite{BookFlanagan} for a detailed derivation) and is summarised here.
The Earth and star are assumed to be at rest in flat space. 
The coordinate components of the photon's four-momentum are not directly observable; instead an observer on Earth measures the \emph{tetrad} components of the photon's four-momentum and from these is able to deduce the star's astrometric position (the unit vector $\vec{n}$), and the frequency of the starlight.
A monochromatic plane-fronted GW, from the direction of the unit vector $\vec{q}$, gives the metric perturbation\footnote{When working with astrometry it is natural to define the sky position of the GW source, $\vec{q}$; this differs from the usual PTA convention where the GW propagation direction, $\vec{\Omega}\!=\!-\vec{q}$, is used.}
\begin{align}\label{eq:metricperturbation} h_{\mu\nu}(t,\vec{x})=\Re\big\{H_{\mu\nu}\exp(\rmi k_{\rho}x^{\rho})\big\}\,, \end{align}
where $H_{\mu\nu}$ are small complex constants satisfying the usual transverse-traceless gauge conditions and the wavevector, ${k^{\mu}\!=\!(\omega, -\omega\vec{q})}$, is null.

The observed photons follow null geodesics from the star to the Earth; integrating the geodesic equations gives the change in the \emph{coordinate} components of the photon four-momentum. The GW also changes the Earth-bound observer's tetrad, this may be calculated by integrating the parallel transport equations along the worldline of the Earth. Combining these results gives the change in the \emph{tetrad} components of the photon four-momentum, and hence the measured frequency and astrometric position.

The frequency perturbation is described by the redshift, defined as $1+z\equiv \Omega_{\textrm{emit}}/\Omega_{\textrm{obs}}$, which is given by
\begin{equation} z  = \frac{n^{i}n^{j}}{2(1-\vec{q}\cdot\vec{n})}\left[h_{ij}(\textrm{E})\!-\!h_{ij}(\textrm{S})\right] \label{eq:PTA_redshift}\,;\end{equation}
this result is the foundation of PTA efforts to detect GWs \cite{1970Natur.227..157K,1975GReGr...6..439E}.
The redshift depends (anti)symmetrically on the metric perturbations at the ``emission'' and ``absorption'' events at the star and Earth respectively ($h_{ij}(\textrm{S})$ and $h_{ij}(\textrm{E})$). This symmetry arises from the endpoints of the integral along the null geodesic from the star to the Earth. This redshift (when applied to a pulsar) can be integrated to give the timing residual signal searched for by PTAs.

The astrometric perturbation also depends on the metric perturbations at the star and at the Earth, although not symmetrically. This loss of symmetry arises from perturbations to the spatial vectors in the observer's tetrad which depend only on the metric at the Earth. The full expression for the astrometric deflection is lengthy, however it simplifies considerably in the limit where the star is many gravitational wavelengths away from Earth \cite{1996ApJ...465..566P};
\begin{align}\label{eq:AstroDefFinal} 
\delta n_{i} = \frac{n_{i}-q_{i}}{2(1-\vec{q}\cdot\vec{n})}h_{jk}(\textrm{E})n^{\Jmath}n^{k} -\frac{1}{2}h_{ij}(\textrm{E})n^{j}\,.
\end{align}

\begin{figure}[t]
\includegraphics[trim={0 1cm 0 0.5cm},width=0.41\textwidth]{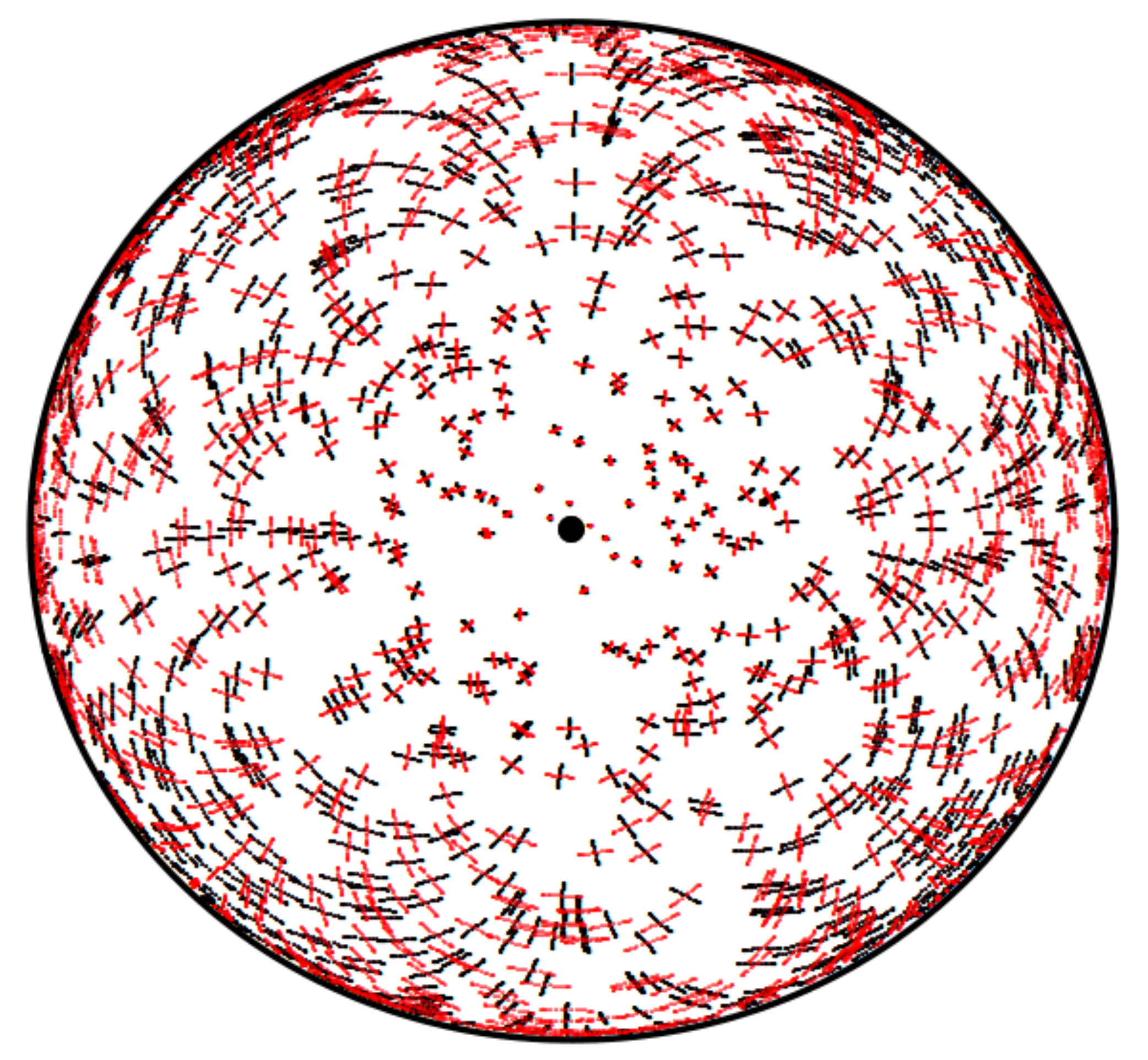}
\caption{Orthographic projection of the Northern hemisphere with $10^{3}$ stars. A GW from the North pole (black dot) causes stars to oscillate at the GW frequency. The black (red) lines show movement tracks for a linearly plus (cross) polarised GW. For clarity, the GW has an unphysically large strain amplitude of $A\!=\!0.1$. The four-fold rotational symmetry of the transverse--traceless GWs is clearly imprinted on the sky.\label{fig:OrthographicProj}}
\end{figure}

In this limit the astrometric deflection depends only on the ``Earth term''. The ``star term'' (or ``pulsar term'') is also sometimes dropped in PTA searches for individually resolvable sources, but for a different reason. Because each pulsar is at a different (generally poorly constrained) distance from Earth the ``pulsar terms'' are all at different frequencies and phases and may be treated as an effective source of noise. Recent searches have tended to include the ``pulsar terms'' (e.g.\ see recent published searches for individual supermassive black hole binaries from the three main PTAs \cite{0004-637X-794-2-141,2016MNRAS.455.1665B,Zhu11112014}, and references therein) which has the benefit of increasing the observed signal-to-noise at the expense of fitting for the distance to each pulsar (for a discussion of the benefits of including the pulsar term see, e.g.\ \cite{2016MNRAS.461.1317Z}).

Gaia's sensitivity to GWs comes from the large number of stars it observes. Stars are typically separated by many gravitational wavelengths, therefore each ``star term'' will be different (as well as being suppressed by the distance to the star) whereas the ``Earth term'' is dominant and common to all stars. It is this common ``Earth term'' that Gaia aims to detect. Including the ``star term'' marginally increases the signal-to-noise ratio for the closest few stars but makes a negligible difference for the majority (e.g. a GW with wavelength $\lambda\!=\!10^{16}\,\textrm{m}$ deflecting a typical star at $d\!=\!10\,\textrm{kpc}$ gives a ``star term'' suppressed by $\lambda/d\!\approx\!10^{5}$). Shown in Fig.~\ref{fig:OrthographicProj} are ``Earth term'' astrometric deflection patterns for a field of distant stars. If the star terms were included they would appear as a random noise superposed on the regular ``Earth term'' pattern with an amplitude greatly suppressed by the number of GW wavelengths to the source.

\medskip
\noindent{\bf \em Data analysis~--~}
This section describes how to search for a monochromatic GW in an astrometric data set. The likely astrophysical source of such a GW is a circular supermassive binary black hole binary with total mass in the range $(10^{7}\textrm{--}10^{10})M_{\odot}$. Such systems spend most of their lifetime in the relatively weak gravitational field where they can be safely assumed to be non-evolving over the observation period
\footnote{For a binary to be considered monochromatic for Gaia analysis, the timescale, $\tau$, over which the GW frequency, $f_{\textrm{GW}}$, evolves must exceed the mission lifetime of $\approx\!10\,\textrm{years}$. This timescale can be estimated via $\tau\!\approx\!f_{\textrm{GW}}/\dot{f}_{\textrm{GW}}$ using leading order post-Newtonian expressions (see, e.g.\ \cite{2007arXiv0709.4682B}). All binaries satisfy $\tau\!>\!10\,\textrm{years}$ up to $\approx\!3.5\,\textrm{years}$ before merger, independent of the component masses. In contrast, these systems cannot always be considered monochromatic for PTA analysis because the ``pulsar terms'' provide snapshops of the $f_{\textrm{GW}}$ at widely seperated times allowing the frequency evolution to be measured (see, e.g.\ \cite{PhysRevD.90.104028}).}.
Points on the sky are denoted as $\vec{n}$, and vectors tangent to the sky are denoted as $\mathbf{h}$. For small vectors $\left|\mathbf{h}\right|\!\ll\!1$, e.g.\ the GW astrometric deflection, the sum ${\vec{n}\,'\!=\!\vec{n}\!+\!\mathbf{h}}$ gives a nearby point on the sphere.

The metric perturbation for a plane, monochromatic GW may be written as
\begin{equation}\label{eq:AstrometricSignal} h_{ij}\left(\overline{\Psi}\right)\!=\!\left(A_{+}H^{+}_{ij}(\vec{q})e^{\rmi\phi_{+}}\!+\!A_{\times}H^{\times}_{ij}(\vec{q})e^{\rmi\phi_{\times}}\right)e^{2\pi\rmi f  t} \,, \end{equation} 
where $H^{+}_{ij},\,H^{\times}_{ij}$ are the usual GW basis tensors, and $\overline{\Psi}$ is a 7-dimensional parameter vector: two amplitudes $A_{+},\,A_{\times}$, two phases $\phi_{+},\,\phi_{\times}$, the GW frequency $f$, and two angles describing the direction $\vec{q}$ to the GW source.

The data set, $\mathcal{S}$, consists of $N$ separate astrometric measurements of $M$ stars. The different stars (and measurements) are indexed by $I$ (and $J$). The observationss are made at times $t_{J}$ (for simplicity the $t_{J}$ are assumed to be the same for all stars, although this is not required);
\begin{equation} \mathcal{S}=\left\{ \vec{s}_{I,J} | I=1,2,\ldots,M ;\, J=1,2,\ldots,N\right\} \,. \label{eq:astrometric_dataset}\end{equation}

Each individual measurement is a combination of the background star position at that time ($\vec{n}_{I}(t_{J})$), noise in the instrument ($\mathbf{r}_{I,J}$), and (possibly) a GW;
\begin{equation} \vec{s}_{I,J} = \vec{n}_{I}(t_{J}) + \mathbf{r}_{I,J} + \mathbf{h}\left(\overline{\Psi}; \vec{n}_{I}(t_{J}),t_{J}\right) \,.\end{equation}

The background star positions vary due to the star's proper motion. For each star the function $\vec{n}_{I}(t_{J})$ is modelled as a quadratic, $\vec{\underline{n}}_{I}(t_{J})$ and subtracted from the data;
\begin{equation} \label{eq:sub_fitting_model} \mathbf{s}_{I,J} = \vec{s}_{I,J}-\vec{\underline{n}}_{I}(t_{J})\,. \end{equation}
Thereby the background positions, proper motions, and accelerations are fit out of the data. This is the astrometric equivalent of the pulsar \emph{timing model} and sets the low frequency sensitivity \cite{2015CQGra..32e5004M}. The position model can be marginalised over (see \cite{2009MNRAS.395.1005V} in the PTA context) however, here the maximum likelihood model parameters are used.

For simplicity the noise in each measurement is assumed to be identical and independent ($\sigma\!\equiv\!\sigma_{I,J}$, again, this is not required),
\begin{equation} \textrm{E}\left[\mathbf{r}_{I,J}\cdot\mathbf{r}_{I',J'}\right] = \sigma^{2}\delta_{II'}\delta_{JJ'}\,. \label{eq:noise}\end{equation}

The likelihood of observing the dataset $\mathcal{S}$ given the GW parameters $\overline{\Psi}$, assuming the star's position and motion have been correctly fit for and under the noise assumptions described, may be written as
\begin{equation}\label{eq:Likelihood} P\left(\mathcal{S}|\overline{\Psi}\right) \!\propto\! \exp\!\left(\sum_{I = 1}^{M}\!\sum_{J = 1}^{N}\!\frac{\textrm{--}\left|\mathbf{s}_{I,J}\textrm{--}\mathbf{h}\left(\overline{\Psi}; \tilde{\mathbf{n}}_{I}(t_{J}), t_{J}\right)\right|^{2}}{2\sigma^{2}}\right), \end{equation}
where $\left| \cdot \right|$ denotes the norm of a vector on the sphere. The posterior probability follows from Bayes' theorem,
\begin{align} \label{eq:BayesThrm} P\left(\overline{\Psi}|\mathcal{S}\right)= \frac{\Pi\left(\overline{\Psi}\right)P\left(\mathcal{S}|\overline{\Psi}\right)}{\mathcal{Z}_{\textrm{signal}}}\,, \end{align}
where $\Pi\left(\overline{\Psi}\right)$ is the prior. Throughout this letter uniform periodic priors for the phase angles $\phi_{+},\,\phi_{\times}$, uniform in log priors for the amplitudes $A_{+},\,A_{\times}$, uniform in log prior for the frequency in the range $f\!\sim\!\mathcal{U}\left[1/T,N/2T\right]$, and a uniform prior on the sphere for $\vec{q}$ are used. 

The Bayesian signal evidence normalises the distribution in Eq.~\ref{eq:BayesThrm} and is given by 
\begin{align} \mathcal{Z}_{\textrm{signal}}=\int\textrm{d}\overline{\Psi}\; \pi\left(\overline{\Psi}\right)P\left(\mathcal{S}|\overline{\Psi}\right)\,. \label{eq:signal_evidence}\end{align}
The noise evidence $\mathcal{Z}_{\textrm{noise}}$, is simply given by the likelihood in Eq.~\ref{eq:Likelihood} evaluated with no GW signal. The Bayes' factor $\mathcal{B}\!\equiv\!\mathcal{Z}_{\textrm{signal}}/\mathcal{Z}_{\textrm{noise}}$ is used as a detection statistic; it is assumed that any signal with $\mathcal{B}_{\textrm{threshold}}\!=\!10^{1.5}$ can be confidently detected. This is generally a conservative choice, and corresponds to Jeffrey's \cite{Jeffreys} criterion for detection with ``very strong'' evidence; the precise detection threshold is problem specific and will depend of the details of the final Gaia data release as they become known.

The \textsc{MultiNest} \cite{2008MNRAS.384..449F} implementation of nested sampling \cite{JohnSkilling} was used to simultaneously sample the posterior (Eq.~\ref{eq:BayesThrm}) and evaluate the Bayesian evidence (Eq.~\ref{eq:signal_evidence}).

\begin{figure*}[t]
\includegraphics[trim={0 0.5cm 0 0.1cm},width=0.99\textwidth]{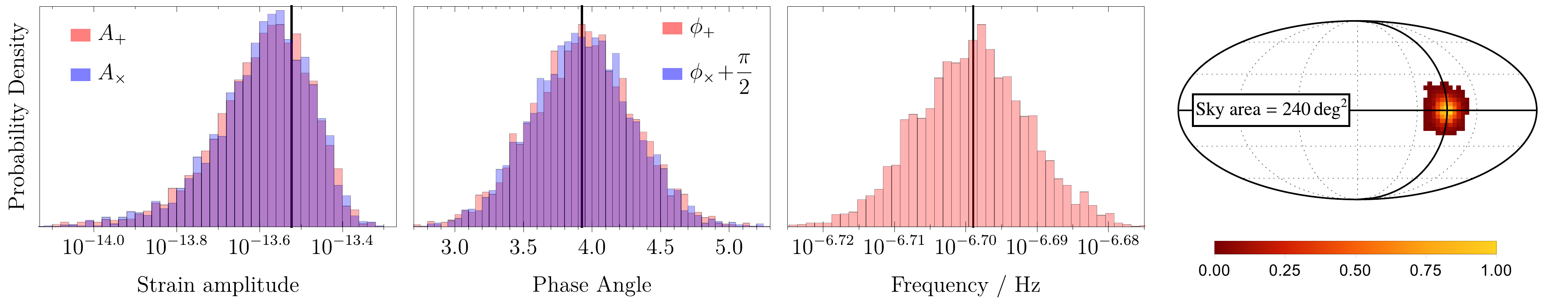}
\caption{1--dimensional marginalised posteriors on $\overline{\Psi}$ (black lines indicate injected values). The injected GW was circularly polarised (i.e $\phi_{+}\!-\phi_{\times}\!=\!\pi/2$) so the $\phi_{\times}$ posterior is shifted such that it overlaps with $\phi_{+}$. The Mollweide sky map is shown with the area of the $68\%$ credible region given. \label{fig:posterior}} 
\end{figure*}

A mock Gaia data set was constructed consisting of $M\!=\!10^{5}$ stars (approximately a factor of $10^{4}$ less than the full Gaia catalog for computational necessity) each measured $N\!=\!75$ times evenly spaced over a $T\!=\!5\,\textrm{year}$ mission (the effect of non-uniformity in the Gaia sampling function is explored below). The simulated noise in each measurement was $\sigma\!=\!100\,\mu\textrm{as}/\sqrt{10^{4}}$, reflecting an estimate of the errors in each measurement in Gaia's final data release and the reduced number of stars (the validity of this scaling and our ability to achieve the compression is established below). For each star the position model $\vec{\underline{n}}_{I}(t_{J})$ was fitted, and subtracted according to Eq.~\ref{eq:sub_fitting_model}. 

The sensitivity of the mock data is largely determined by $N$, $M$, $T$ and $\sigma$; the values of $N$, $M$, $T$ used are pessimistic estimates for the final Gaia data release while the value of $\sigma$ is slightly optimistic. In particular, Gaia errors vary strongly with magnitude \cite{2010IAUS..261..306M}; a simple estimate of the appropriate error in each measurement derived by averaging over the full magnitude range, using fits to the $G$--magnitude distribution \cite{GmagHist}, yielded a conservative estimate of $200\,\mu\textrm{as}$. Thus, the data set used here reflects our current best guess of Gaia's ultimate sensitivity but should be updated with more accurate estimates following future Gaia data releases.

A GW from a high mass, non-spinning binary was injected into this data set; black holes with masses $m_{1}\!=\!m_{2}\!=\!5\times10^{8}M_{\odot}$ on a circular orbit of radius $1500\,\textrm{au}$ at a distance of $20\,\textrm{Mpc}$ (orientated with the angular momentum along the line-of-sight) give a circularly polarised GW with frequency $2\pi f\!=\!2\times10^{-7}\textrm{Hz}$ and amplitude $A_{+}\!=\!A_{\times}\!=\!3\times10^{-14}$. The GW was confidently recovered with ${\mathcal{B}\!=\!10^{4.2}\!>\!\mathcal{B}_{\textrm{threshold}}}$ and the 1--dimensional marginalised posterior distributions are shown in Fig.~\ref{fig:posterior}.

\medskip
\noindent{\bf \em Compressing the GAIA dataset~--~}
Calculations with $M\!=\!10^{5}$ stars take days to run; the full Gaia data set consisting of $M\!>\!10^{9}$ stars is impractically large to search using the Bayesian techniques described. 
The need for efficient compression will be even greater when performing an astrometric search for a stochastic background of GWs because the likelihood depends on the inverse of a $M\!\times\!M$ correlation matrix \cite{BookFlanagan} (compression for stochastic searches will be addressed in a future publication). In this section it is demonstrated how the data can be greatly compressed with little loss in sensitivity.

A small number $\tilde{M}(\ll\!M)$ of points on the sky are selected; these are called \emph{virtual stars}. Each virtual star defines a \emph{Voronoi cell} \cite{Sack:2000:HCG:337150} consisting of the points closest to that virtual star. 
Each real star is identified with the nearest virtual star.
Virtual stars are indexed by $\tilde{I}\!=\!1,2,\ldots,\tilde{M}$ and the Voronoi cells are denoted $\mathcal{V}_{\tilde{I}}$.

The large astrometric data set is compressed onto a smaller \emph{virtual data set} (quantities associated with the virtual data set are denoted with a tilde). All of the astrometric deflections in a given time interval for stars in a given Voronoi cell are averaged;
\begin{equation}\label{eq:compression} \tilde{\mathbf{s}}_{\tilde{I},J} = \frac{1}{\left|\mathcal{V}_{\tilde{I}}\right|}\sum_{I\in\mathcal{V}_{\tilde{I}}}\mathbf{s}_{I,J} \,,\quad\frac{1}{\tilde{\sigma}_{\tilde{I},J}^{2}} = \sum_{I\in\mathcal{V}_{\tilde{I}}}\frac{1}{\sigma_{I,J}^{2}} \,, \end{equation} 
where $|\mathcal{V}_{\tilde{I}}|$ denotes the number of real stars in $\mathcal{V}_{\tilde{I}}$. The virtual data ${\tilde{\mathcal{S}}\!=\!\small\{ \tilde{\mathbf{s}}_{\tilde{I},J} | \tilde{I}\!=\!1,\ldots,\tilde{M} ;\, J\!=\!1,\ldots,N\small\}}$ (c.f.\ Eq.~\ref{eq:astrometric_dataset}) may be analysed using the techniques described above for the original data, $\mathcal{S}$.

\begin{figure}[b]
\includegraphics[trim={0 0.5cm 0 0.3cm},width=0.4\textwidth]{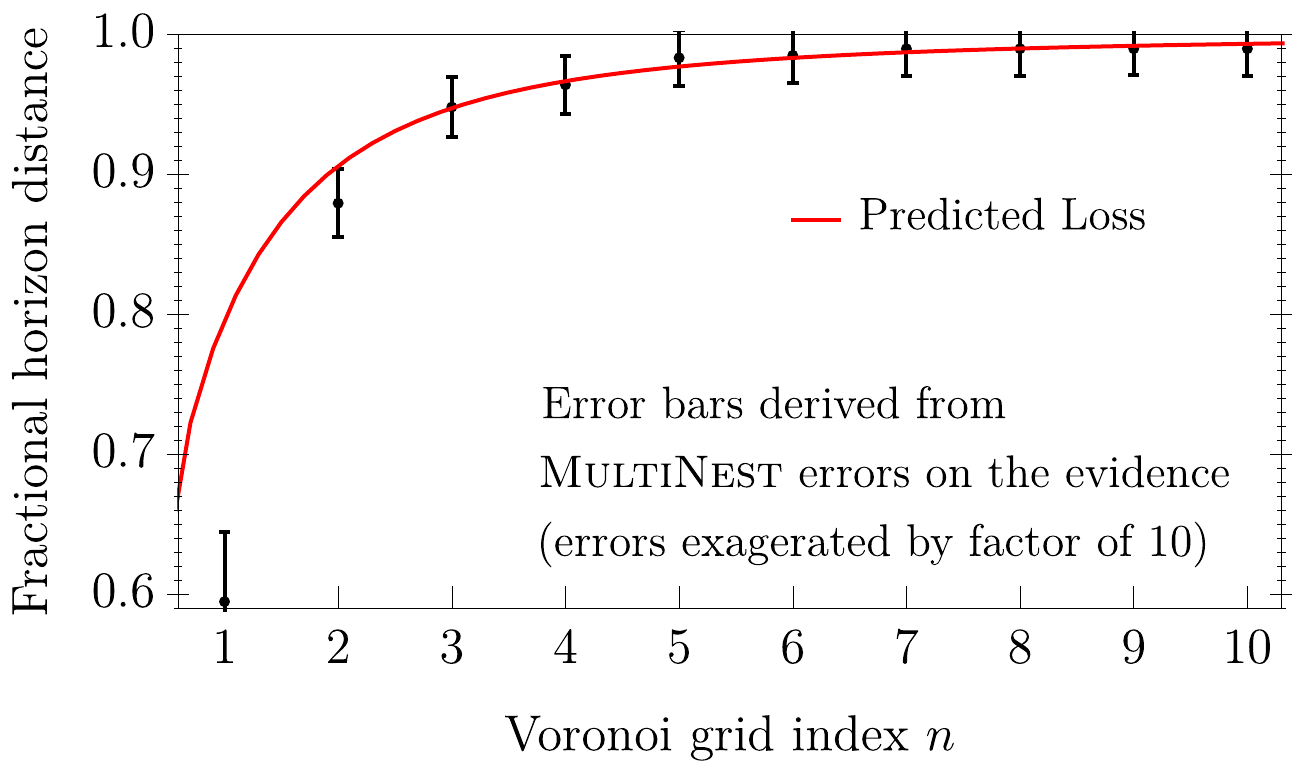}
\caption{The horizon distance (relative to that for the uncompressed data) is reduced during compression onto Voronoi grid $n\!=\!1,2,\ldots,10$. Shown in red is an estimate of the loss obtained by considering the angle between astrometric deflections for stars in the same Voronoi cell.\label{fig:compression_loss}}
\end{figure}

This compression would be lossless if (i) the noise was independent as described by Eq.~\ref{eq:noise}, and (ii) the astrometric deflections of all the stars in a Voronoi cell were parallel. The deflections vary smoothly across the sky (see Fig.~\ref{fig:OrthographicProj}) so as $\tilde{M}$ is increased condition (ii) becomes satisfied. In fact, for a given $\tilde{M}$ the sensitivity loss can be estimated by considering the angle between deflections of stars in the same Voronoi cell (see Fig.~\ref{fig:compression_loss}). 

While condition (i) cannot be expected to hold perfectly for Gaia, correlations are not expected to significantly degrade the sensitivity. Correlations in time will be mitigated against by the fact that between taking successive measurements of a star the spacecraft rotates into a different orientation and the starlight strikes a different part of the CCD array. By contrast, ``red noise'' temporal correlations do limit the sensitivity of PTAs. Spatial correlations across the sky do exist but only at the percent level (correlations of 3\% for colocated stars, dropping to 0\% for stars seperated by $0.7\degree$). As the Gaia mission proceeds correlations are expected to decrease, with final mission products being essentially uncorrelated compared to their random errors \cite{2012A&A...543A..15H}. Thus for this first analysis we do not consider such correlated errors.

The locations of the virtual stars may be freely specified, e.g.\ they could be randomly placed on the sky. Here they are taken to be at the midpoints of the faces of certain polyhedra. The base polyhedron was taken to be an icosahedron (the resulting Voronoi cells are called ``grid 1''). Successive polyhedra were formed by constructing geodesic domes from the icosahedron --- subdividing great circle arcs between vertices into $n \! = \! 2, 3, \dots$ smaller arcs, and then constructing $n^{2}$ triangles on each face.  The midpoints of the faces of the resulting polyhedra give a set of virtual stars and the resulting Voronoi cells are called ``grid $n$''. The $n^{\textrm{th}}$ Voronoi grid has $\tilde{M}\!=\!20 \times n^{2}$ virtual stars; grids up to $n\!=\!10$ were used. A controllable level of compression can be achieved by varying $n$.

The mock data described above was compressed onto each of the grids $n\!=\!10, 9,\ldots, 1$ and the virtual data sets searched as before. The Bayes' factor recovered from smaller grids is reduced because stars in the larger Voronoi cells have astrometric deflections which are not parallel and partially cancel each other out in the compression (Eq.~\ref{eq:compression}). This lower Bayes' factor reduces the maximum distance at which the source can be detected; this reduction in \emph{horizon distance} is shown in Fig.~\ref{fig:compression_loss}. The compression loss is independent of the number of real stars. Provided grids with $n\!\geq\!7$ are used the sensitivity loss is less than $1\%$. The $n\!=\!7$ grid contains $\tilde{M}\!=\!980$ virtual stars; therefore the full Gaia data containing $M\!>\!10^{9}$ stars can be compressed onto the $n\!=\!7$ grid (a compression factor of $10^{9}/980\!\approx\!10^{6}$) with a sensitivity loss below $1\%$. The averaging in Eq.~\ref{eq:compression} gives these impressive compressions because of the smooth, large angle (approximately quadrupolar) pattern in Fig.~\ref{fig:OrthographicProj}.

\medskip
\noindent{\bf \em GAIA's frequency sensitivity~--~}
In this section the frequency dependence of Gaia's sensitivity is quantified, along with the effect of nonuniform time sampling. A large number of mock data sets, similar to that used previously were constructed. The astrometric position of each star was measured $N\!=\!75$ times over a $T\!=\!5\,\textrm{year}$ period; some data sets were constructed assuming uniform time sampling ($T_{0}$), and some using realistic Gaia samplings constructed using Gaia tools (\url{https://gaia.esac.esa.int/gost/}) applied to three points on the sky chosen to give a representation of the variability in the Gaia sampling function (these three samplings were labeled $T_{\alpha}$, for $\alpha\!=\!1,2,3$).

\begin{figure}[h]
\includegraphics[trim={0 0.5cm 0 0.3cm},width=0.46\textwidth]{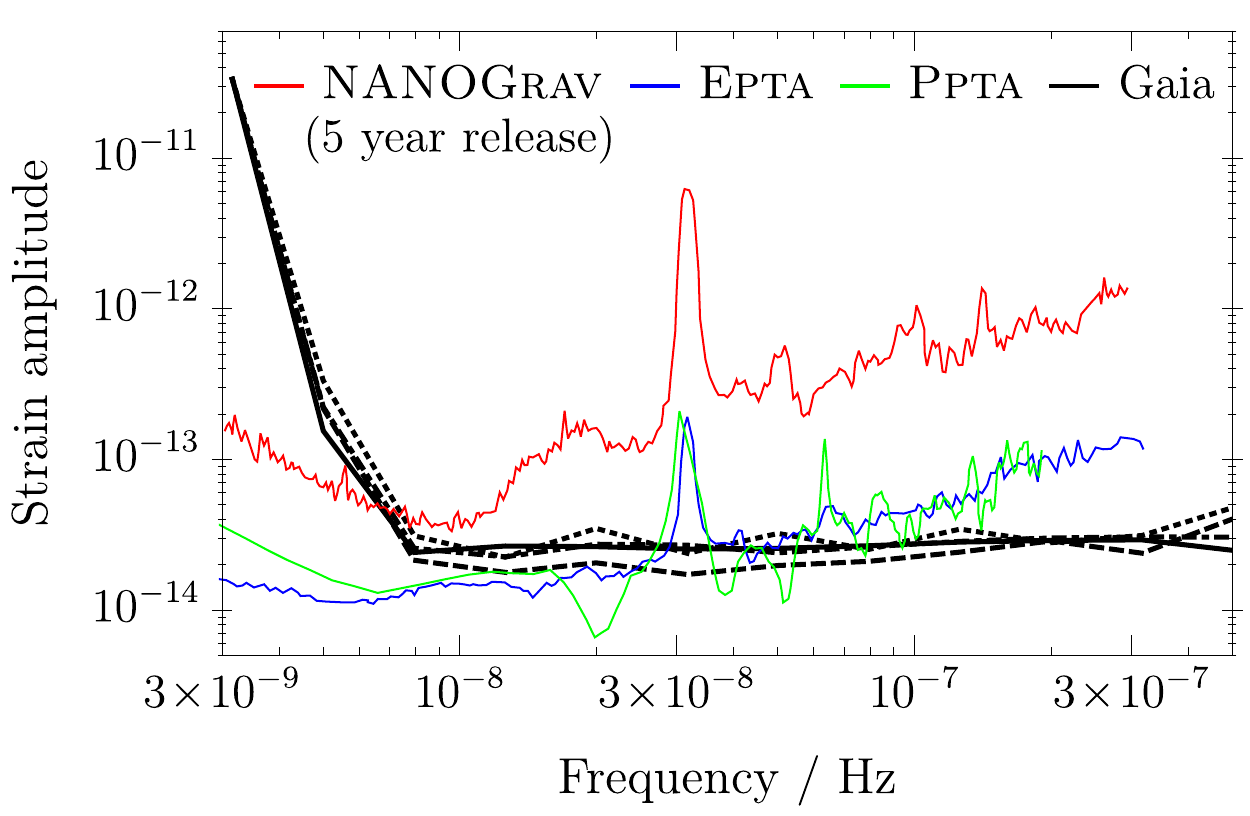}
\caption{The thick black curves show the strain sensitivity of the final Gaia data release using the different time samplings; $T_{0}$ is the solid line, $T_{1}$ is the dotted line, $T_{2}$ is the dashed line, and $T_{3}$ is the dot-dashed line. The four curves are very similar. For comparison the thin coloured lines show the $95\%$ upper limits from the three PTA collaborations: {\smallcaps{NanoG}\!rav} (\cite{0004-637X-794-2-141} red), \smallcaps{Epta} (\cite{2016MNRAS.455.1665B} blue) and \smallcaps{Ppta} (\cite{Zhu11112014} green). 
The curves in this plot show different quantities and are only intended to allow for approximate comparissons; the {\smallcaps{NanoG}\!rav} curve is a Bayesian 95\% upper limit, the \smallcaps{Epta} and \smallcaps{Ppta} curves are frequentist 95\% upper limits, while the Gaia curves show the amplitude necessary to achieve a (conservative) threshold Bayes' factor.
It should be noted that the PTA limits plotted are several years old and constraints improve over time; Gaia's sensitivity will not improve further. However, it is clear that, especially at higher frequencies, Gaia promises to provide a useful complement to the existing limits from pulsar timing. \label{fig:gaia_sensitivity}}
\end{figure}

Circularly polarised GWs were injected with different amplitudes and frequencies and the data sets were compressed onto the $n\!=\!10$ Voronoi grid for analysis. For multiple fixed frequencies in the range $(10^{-8.5}\textrm{--}10^{-6})\textrm{Hz}$ several mock injections were used to find the minimum amplitude necessary for detection. The resulting sensitivity curves are shown in Fig.~\ref{fig:gaia_sensitivity} for each of the $T_{\alpha}$, demonstrating that the variability in Gaia's sampling has only a minor effect on its sensitivity to GWs.

The strain sensitivity of Gaia is almost flat above $f\!\gtrsim\!1/T$ (where $T\!=\!5\,\textrm{years}$ is the mission lifetime). This is in sharp contrast to the sensitivity of PTAs which degrade linearly at higher frequencies. This discrepancy comes from the fact that GWs cause a redshift (Eq.~\ref{eq:PTA_redshift}) and PTAs measure the \emph{timing residual} which is the integral of redshift over time. In the frequency domain, integration over time corresponds to division by frequency; this integration suppresses the sensitivity of PTAs for frequencies above $f\!\approx\!1/T$. In contrast, Gaia measures the astrometric deflection which is directly proportional to the GW strain (Eq.~\ref{eq:AstroDefFinal}). This difference in slopes means that it is likely to be at mid to high frequencies, $f\!\gtrsim\!10^{-7.5}\textrm{Hz}$, where Gaia will best complement current PTA efforts.

\medskip
\noindent{\bf \em GAIA's directional sensitivity~--~}
The distribution of stars on the sky is not uniform (as was assumed for simplicity in the previous section), therefore astrometric measurements are not uniformly sensitive to GWs from all directions. In this section the directional dependence of Gaia's GW sensitivity is quantified.

Without loss of generality let the GW source lie on the positive z--axis ($\vec{q}\!=\!\left\{0,0,1\right\}$) and a star lie in the $x$--$z$ plane ($\vec{n}\!=\!\left\{\sin\gamma,0,\cos\gamma\right\}$). Using the general metric perturbation in Eq.~\ref{eq:AstrometricSignal}, the magnitude of the astrometric deflection vector in Eq.~\ref{eq:AstroDefFinal} is given by 
\begin{align}\label{eq:angularresponse} \left| \delta \vec{n} \right| = \frac{1}{2}\sqrt{A_{+}^{2}+A_{\times}^{2}}\sin\gamma \;. \end{align}
The largest deflections occur for stars orthogonal to the GW source direction (i.e.\ $\gamma\!=\!\pi/2$). Therefore, it is expected that Gaia's peak sensitivity will occur orthogonal to regions of high stellar density (i.e.\ the galactic poles).

\begin{figure}[t]
\includegraphics[trim={0 0.5cm 0 0.1cm},width=0.42\textwidth]{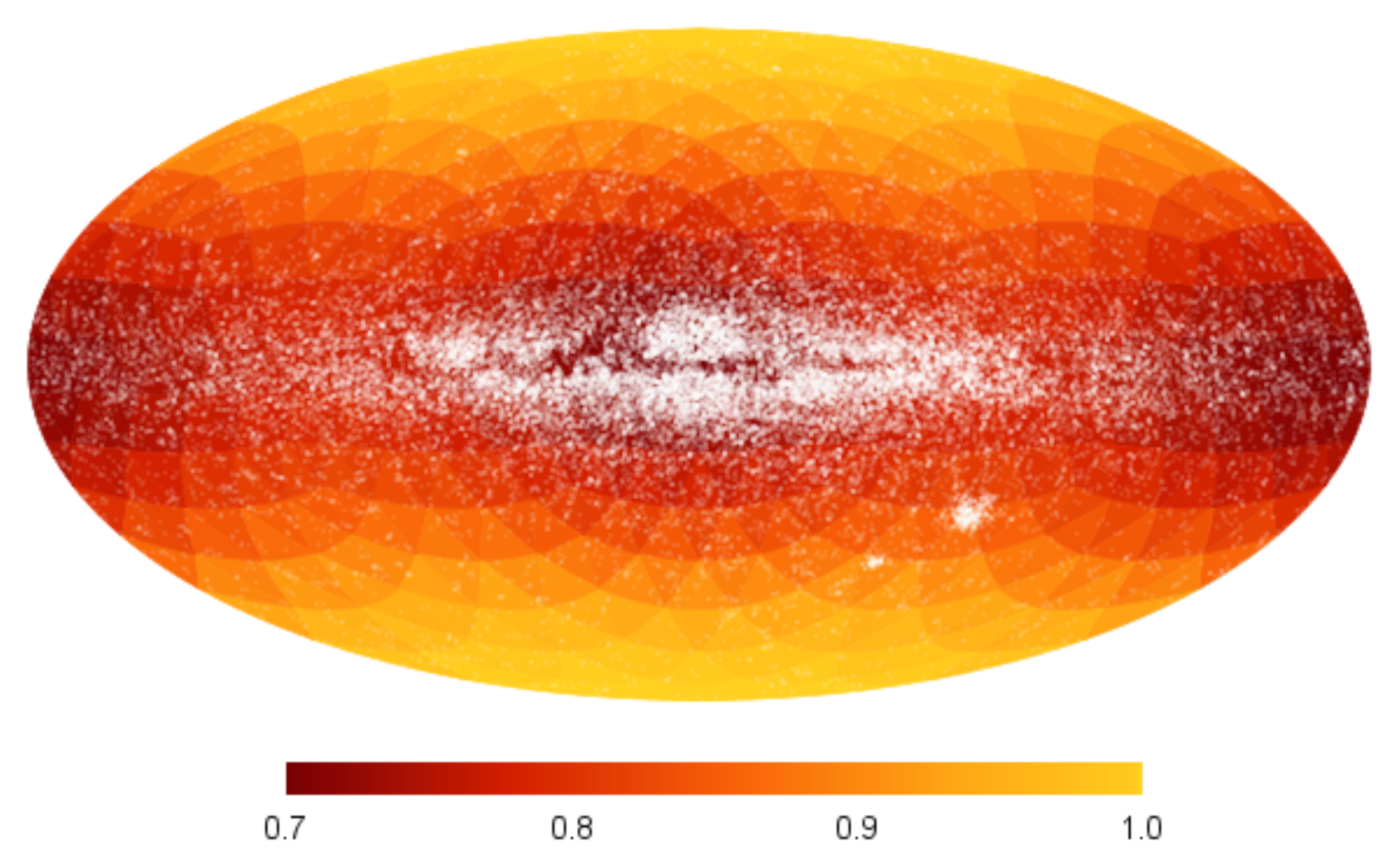}
\caption{The variation in Gaia's sensitivity over the sky, $F(\theta,\phi)$. A sample of 60,000 stars drawn randomly from the Gaia catalogue are shown as white dots. The sensitivity varies by $\sim\!30\%$ across the sky with minima at (and antipodal to) the galactic centre, and maxima at the galactic poles.\label{fig:gaia_SkyMap}}
\end{figure}

Mock data sets were constructed using the ${M\!=\!1.1\times 10^{9}}$ real stars in the first Gaia data release (\url{https://www.cosmos.esa.int/web/gaia/dr1}). The astrometric positions were sampled $N\!=\!75$ times uniformly over a $T\!=\!5\,\textrm{year}$ mission. Into these mock data sets were injected circularly polarised GWs from 500 sky locations. The data were compressed onto the $n\!=\!5$ grid to be efficiently searched. The variation in horizon distance with sky location is plotted in Fig.~\ref{fig:gaia_SkyMap}. 

\medskip
\noindent{\bf \em Conclusions~--~}
GWs cause the apparent astrometric position of distant stars to oscillate with a characteristic pattern (see Fig.~\ref{fig:OrthographicProj} and Eq.~\ref{eq:AstroDefFinal}) on the sky. Gaia is the ideal observatory to make the large number of accurate astrometric measurements necessary to search for low frequency GWs using this effect. This letter summarises recent progress towards a data analysis pipeline to search for GWs in the fast approaching final Gaia data release. It has been shown how a large astrometric data set may be greatly compressed with little loss in sensitivity; this is vital to enable a GW search to be performed. A large number of mock injections have been performed to quantify the sensitivity of Gaia, and establish the accuracy with which GW parameters can be measured.

\medskip
\noindent{\bf \em Acknowledgments~--~} This work has been supported by Grant Nos. MSCA-RISE-2015 690904, STFC ST/L000636/1, ST/H008586/1 and ST/K00333X/1, and BIS ST/J005673/1. Thanks to Jonathan Gair, Stephen Taylor, Michalis Agathos and Ulrich Sperhake for helpful comments on the manuscript. We thank the anonymous referee for comments which improved the manuscript.

\bibliography{bibliography}

\begin{thebibliography}{32}%
\makeatletter
\providecommand \@ifxundefined [1]{%
 \@ifx{#1\undefined}
}%
\providecommand \@ifnum [1]{%
 \ifnum #1\expandafter \@firstoftwo
 \else \expandafter \@secondoftwo
 \fi
}%
\providecommand \@ifx [1]{%
 \ifx #1\expandafter \@firstoftwo
 \else \expandafter \@secondoftwo
 \fi
}%
\providecommand \natexlab [1]{#1}%
\providecommand \enquote  [1]{``#1''}%
\providecommand \bibnamefont  [1]{#1}%
\providecommand \bibfnamefont [1]{#1}%
\providecommand \citenamefont [1]{#1}%
\providecommand \href@noop [0]{\@secondoftwo}%
\providecommand \href [0]{\begingroup \@sanitize@url \@href}%
\providecommand \@href[1]{\@@startlink{#1}\@@href}%
\providecommand \@@href[1]{\endgroup#1\@@endlink}%
\providecommand \@sanitize@url [0]{\catcode `\\12\catcode `\$12\catcode
  `\&12\catcode `\#12\catcode `\^12\catcode `\_12\catcode `\%12\relax}%
\providecommand \@@startlink[1]{}%
\providecommand \@@endlink[0]{}%
\providecommand \url  [0]{\begingroup\@sanitize@url \@url }%
\providecommand \@url [1]{\endgroup\@href {#1}{\urlprefix }}%
\providecommand \urlprefix  [0]{URL }%
\providecommand \Eprint [0]{\href }%
\providecommand \doibase [0]{http://dx.doi.org/}%
\providecommand \selectlanguage [0]{\@gobble}%
\providecommand \bibinfo  [0]{\@secondoftwo}%
\providecommand \bibfield  [0]{\@secondoftwo}%
\providecommand \translation [1]{[#1]}%
\providecommand \BibitemOpen [0]{}%
\providecommand \bibitemStop [0]{}%
\providecommand \bibitemNoStop [0]{.\EOS\space}%
\providecommand \EOS [0]{\spacefactor3000\relax}%
\providecommand \BibitemShut  [1]{\csname bibitem#1\endcsname}%
\let\auto@bib@innerbib\@empty
\bibitem [{\citenamefont {Abbott}\ \emph {et~al.}(2016)\citenamefont {Abbott}
  \emph {et~al.}}]{PhysRevLett.116.061102}%
  \BibitemOpen
  \bibfield  {author} {\bibinfo {author} {\bibfnamefont {B.~P.}\ \bibnamefont
  {Abbott}} \emph {et~al.} (\bibinfo {collaboration} {LIGO Scientific- and
  Virgo Collaborations}),\ }\href {\doibase 10.1103/PhysRevLett.116.061102}
  {\bibfield  {journal} {\bibinfo  {journal} {Phys. Rev. Lett.}\ }\textbf
  {\bibinfo {volume} {116}},\ \bibinfo {pages} {061102} (\bibinfo {year}
  {2016})}\BibitemShut {NoStop}%
\bibitem [{\citenamefont {{Abbott}}\ \emph {et~al.}(2016)\citenamefont
  {{Abbott}} \emph {et~al.}}]{2016ApJ...818L..22A}%
  \BibitemOpen
  \bibfield  {author} {\bibinfo {author} {\bibfnamefont {B.~P.}\ \bibnamefont
  {{Abbott}}} \emph {et~al.} (\bibinfo {collaboration} {LIGO Scientific- and
  Virgo Collaborations}),\ }\href {\doibase 10.3847/2041-8205/818/2/L22}
  {\bibfield  {journal} {\bibinfo  {journal} {\apjl}\ }\textbf {\bibinfo
  {volume} {818}},\ \bibinfo {eid} {L22} (\bibinfo {year} {2016})}\BibitemShut
  {NoStop}%
\bibitem [{\citenamefont {{Seoane}}\ \emph {et~al.}(2013)\citenamefont
  {{Seoane}} \emph {et~al.}}]{2013arXiv1305.5720C}%
  \BibitemOpen
  \bibfield  {author} {\bibinfo {author} {\bibfnamefont {P.~A.}\ \bibnamefont
  {{Seoane}}} \emph {et~al.} (\bibinfo {collaboration} {The eLISA
  Collaboration}),\ }\href@noop {} {\bibfield  {journal} {\bibinfo  {journal}
  {ArXiv e-prints}\ } (\bibinfo {year} {2013})},\ \Eprint
  {http://arxiv.org/abs/1305.5720} {arXiv:1305.5720 [astro-ph.CO]} \BibitemShut
  {NoStop}%
\bibitem [{\citenamefont {{Moore}}\ \emph {et~al.}(2015)\citenamefont {{Moore}}
  \emph {et~al.}}]{2015CQGra..32e5004M}%
  \BibitemOpen
  \bibfield  {author} {\bibinfo {author} {\bibfnamefont {C.~J.}\ \bibnamefont
  {{Moore}}} \emph {et~al.},\ }\href {\doibase 10.1088/0264-9381/32/5/055004}
  {\bibfield  {journal} {\bibinfo  {journal} {Class.\ Quantum Grav.}\ }\textbf
  {\bibinfo {volume} {32}},\ \bibinfo {eid} {055004} (\bibinfo {year}
  {2015})}\BibitemShut {NoStop}%
\bibitem [{\citenamefont {{McLaughlin}}(2013)}]{2013CQGra..30v4008M}%
  \BibitemOpen
  \bibfield  {author} {\bibinfo {author} {\bibfnamefont {M.~A.}\ \bibnamefont
  {{McLaughlin}}},\ }\href {\doibase 10.1088/0264-9381/30/22/224008} {\bibfield
   {journal} {\bibinfo  {journal} {Class.\ Quantum Grav.}\ }\textbf {\bibinfo
  {volume} {30}},\ \bibinfo {eid} {224008} (\bibinfo {year}
  {2013})}\BibitemShut {NoStop}%
\bibitem [{\citenamefont {{Kramer}}\ and\ \citenamefont
  {{Champion}}(2013)}]{2013CQGra..30v4009K}%
  \BibitemOpen
  \bibfield  {author} {\bibinfo {author} {\bibfnamefont {M.}~\bibnamefont
  {{Kramer}}}\ and\ \bibinfo {author} {\bibfnamefont {D.~J.}\ \bibnamefont
  {{Champion}}},\ }\href {\doibase 10.1088/0264-9381/30/22/224009} {\bibfield
  {journal} {\bibinfo  {journal} {Class.\ Quantum Grav.}\ }\textbf {\bibinfo
  {volume} {30}},\ \bibinfo {eid} {224009} (\bibinfo {year}
  {2013})}\BibitemShut {NoStop}%
\bibitem [{\citenamefont {{Hobbs}}(2013)}]{2013CQGra..30v4007H}%
  \BibitemOpen
  \bibfield  {author} {\bibinfo {author} {\bibfnamefont {G.}~\bibnamefont
  {{Hobbs}}},\ }\href {\doibase 10.1088/0264-9381/30/22/224007} {\bibfield
  {journal} {\bibinfo  {journal} {Class.\ Quantum Grav.}\ }\textbf {\bibinfo
  {volume} {30}},\ \bibinfo {eid} {224007} (\bibinfo {year}
  {2013})}\BibitemShut {NoStop}%
\bibitem [{\citenamefont {{Manchester}}\ \emph {et~al.}(2013)\citenamefont
  {{Manchester}} \emph {et~al.}}]{2013CQGra..30v4010M}%
  \BibitemOpen
  \bibfield  {author} {\bibinfo {author} {\bibfnamefont {R.~N.}\ \bibnamefont
  {{Manchester}}} \emph {et~al.},\ }\href {\doibase
  10.1088/0264-9381/30/22/224010} {\bibfield  {journal} {\bibinfo  {journal}
  {Class.\ Quantum Grav.}\ }\textbf {\bibinfo {volume} {30}},\ \bibinfo {eid}
  {224010} (\bibinfo {year} {2013})}\BibitemShut {NoStop}%
\bibitem [{\citenamefont {{Prusti, T.}}\ \emph {et~al.}(2016)\citenamefont
  {{Prusti, T.}} \emph {et~al.}}]{refId0Gaia}%
  \BibitemOpen
  \bibfield  {author} {\bibinfo {author} {\bibnamefont {{Prusti, T.}}} \emph
  {et~al.},\ }\href@noop {} {\bibfield  {journal} {\bibinfo  {journal} {A\&A}\
  }\textbf {\bibinfo {volume} {595}},\ \bibinfo {pages} {A1} (\bibinfo {year}
  {2016})}\BibitemShut {NoStop}%
\bibitem [{\citenamefont {{Vilenkin}}(1981)}]{1981PhRvD..24.2082V}%
  \BibitemOpen
  \bibfield  {author} {\bibinfo {author} {\bibfnamefont {A.}~\bibnamefont
  {{Vilenkin}}},\ }\href {\doibase 10.1103/PhysRevD.24.2082} {\bibfield
  {journal} {\bibinfo  {journal} {\prd}\ }\textbf {\bibinfo {volume} {24}},\
  \bibinfo {pages} {2082} (\bibinfo {year} {1981})}\BibitemShut {NoStop}%
\bibitem [{\citenamefont {{Grishchuk}}(1976)}]{1976PZETF..23..326G}%
  \BibitemOpen
  \bibfield  {author} {\bibinfo {author} {\bibfnamefont {L.~P.}\ \bibnamefont
  {{Grishchuk}}},\ }\href@noop {} {\bibfield  {journal} {\bibinfo  {journal}
  {Pisma v Zhurnal Eksperimentalnoi i Teoreticheskoi Fiziki}\ }\textbf
  {\bibinfo {volume} {23}},\ \bibinfo {pages} {326} (\bibinfo {year}
  {1976})}\BibitemShut {NoStop}%
\bibitem [{\citenamefont {{Wang}}\ \emph {et~al.}(2015)\citenamefont {{Wang}}
  \emph {et~al.}}]{2015MNRAS.446.1657W}%
  \BibitemOpen
  \bibfield  {author} {\bibinfo {author} {\bibfnamefont {J.~B.}\ \bibnamefont
  {{Wang}}} \emph {et~al.},\ }\href {\doibase 10.1093/mnras/stu2137} {\bibfield
   {journal} {\bibinfo  {journal} {\mnras}\ }\textbf {\bibinfo {volume}
  {446}},\ \bibinfo {pages} {1657} (\bibinfo {year} {2015})}\BibitemShut
  {NoStop}%
\bibitem [{\citenamefont {{Arzoumanian}}\ \emph {et~al.}(2015)\citenamefont
  {{Arzoumanian}} \emph {et~al.}}]{2015ApJ...810..150A}%
  \BibitemOpen
  \bibfield  {author} {\bibinfo {author} {\bibfnamefont {Z.}~\bibnamefont
  {{Arzoumanian}}} \emph {et~al.},\ }\href {\doibase
  10.1088/0004-637X/810/2/150} {\bibfield  {journal} {\bibinfo  {journal}
  {\apj}\ }\textbf {\bibinfo {volume} {810}},\ \bibinfo {eid} {150} (\bibinfo
  {year} {2015})}\BibitemShut {NoStop}%
\bibitem [{\citenamefont {{Braginsky}}\ \emph {et~al.}(1990)\citenamefont
  {{Braginsky}}, \citenamefont {{Kardashev}}, \citenamefont {{Polnarev}},\ and\
  \citenamefont {{Novikov}}}]{1990NCimB.105.1141B}%
  \BibitemOpen
  \bibfield  {author} {\bibinfo {author} {\bibfnamefont {V.~B.}\ \bibnamefont
  {{Braginsky}}}, \bibinfo {author} {\bibfnamefont {N.~S.}\ \bibnamefont
  {{Kardashev}}}, \bibinfo {author} {\bibfnamefont {A.~G.}\ \bibnamefont
  {{Polnarev}}}, \ and\ \bibinfo {author} {\bibfnamefont {I.~D.}\ \bibnamefont
  {{Novikov}}},\ }\href@noop {} {\bibfield  {journal} {\bibinfo  {journal}
  {Nuovo Cimento B Serie}\ }\textbf {\bibinfo {volume} {105}},\ \bibinfo
  {pages} {1141} (\bibinfo {year} {1990})}\BibitemShut {NoStop}%
\bibitem [{\citenamefont {{Pyne}}\ \emph {et~al.}(1996)\citenamefont {{Pyne}}
  \emph {et~al.}}]{1996ApJ...465..566P}%
  \BibitemOpen
  \bibfield  {author} {\bibinfo {author} {\bibfnamefont {T.}~\bibnamefont
  {{Pyne}}} \emph {et~al.},\ }\href {\doibase 10.1086/177443} {\bibfield
  {journal} {\bibinfo  {journal} {\apj}\ }\textbf {\bibinfo {volume} {465}},\
  \bibinfo {pages} {566} (\bibinfo {year} {1996})}\BibitemShut {NoStop}%
\bibitem [{\citenamefont {{Book}}\ and\ \citenamefont
  {{Flanagan}}(2011)}]{BookFlanagan}%
  \BibitemOpen
  \bibfield  {author} {\bibinfo {author} {\bibfnamefont {L.~G.}\ \bibnamefont
  {{Book}}}\ and\ \bibinfo {author} {\bibfnamefont {{\'E}.~{\'E}.}\
  \bibnamefont {{Flanagan}}},\ }\href {\doibase 10.1103/PhysRevD.83.024024}
  {\bibfield  {journal} {\bibinfo  {journal} {Phys. Rev. D}\ }\textbf {\bibinfo
  {volume} {83}},\ \bibinfo {eid} {024024} (\bibinfo {year}
  {2011})}\BibitemShut {NoStop}%
\bibitem [{\citenamefont {{Kaufmann}}(1970)}]{1970Natur.227..157K}%
  \BibitemOpen
  \bibfield  {author} {\bibinfo {author} {\bibfnamefont {W.~J.}\ \bibnamefont
  {{Kaufmann}}},\ }\href {\doibase 10.1038/227157a0} {\bibfield  {journal}
  {\bibinfo  {journal} {\nat}\ }\textbf {\bibinfo {volume} {227}},\ \bibinfo
  {pages} {157} (\bibinfo {year} {1970})}\BibitemShut {NoStop}%
\bibitem [{\citenamefont {{Estabrook}}\ and\ \citenamefont
  {{Wahlquist}}(1975)}]{1975GReGr...6..439E}%
  \BibitemOpen
  \bibfield  {author} {\bibinfo {author} {\bibfnamefont {F.~B.}\ \bibnamefont
  {{Estabrook}}}\ and\ \bibinfo {author} {\bibfnamefont {H.~D.}\ \bibnamefont
  {{Wahlquist}}},\ }\href {\doibase 10.1007/BF00762449} {\bibfield  {journal}
  {\bibinfo  {journal} {Gen.\ Rel.\ Gravit.}\ }\textbf {\bibinfo {volume}
  {6}},\ \bibinfo {pages} {439} (\bibinfo {year} {1975})}\BibitemShut {NoStop}%
\bibitem [{\citenamefont {Arzoumanian}\ \emph {et~al.}(2014)\citenamefont
  {Arzoumanian} \emph {et~al.}}]{0004-637X-794-2-141}%
  \BibitemOpen
  \bibfield  {author} {\bibinfo {author} {\bibfnamefont {Z.}~\bibnamefont
  {Arzoumanian}} \emph {et~al.},\ }\href
  {http://stacks.iop.org/0004-637X/794/i=2/a=141} {\bibfield  {journal}
  {\bibinfo  {journal} {\apj}\ }\textbf {\bibinfo {volume} {794}},\ \bibinfo
  {pages} {141} (\bibinfo {year} {2014})}\BibitemShut {NoStop}%
\bibitem [{\citenamefont {{Babak}}\ \emph {et~al.}(2016)\citenamefont {{Babak}}
  \emph {et~al.}}]{2016MNRAS.455.1665B}%
  \BibitemOpen
  \bibfield  {author} {\bibinfo {author} {\bibfnamefont {S.}~\bibnamefont
  {{Babak}}} \emph {et~al.},\ }\href {\doibase 10.1093/mnras/stv2092}
  {\bibfield  {journal} {\bibinfo  {journal} {\mnras}\ }\textbf {\bibinfo
  {volume} {455}},\ \bibinfo {pages} {1665} (\bibinfo {year}
  {2016})}\BibitemShut {NoStop}%
\bibitem [{\citenamefont {Zhu}\ \emph {et~al.}(2014)\citenamefont {Zhu} \emph
  {et~al.}}]{Zhu11112014}%
  \BibitemOpen
  \bibfield  {author} {\bibinfo {author} {\bibfnamefont {X.-J.}\ \bibnamefont
  {Zhu}} \emph {et~al.},\ }\href {\doibase 10.1093/mnras/stu1717} {\bibfield
  {journal} {\bibinfo  {journal} {\mnras}\ }\textbf {\bibinfo {volume} {444}},\
  \bibinfo {pages} {3709} (\bibinfo {year} {2014})}\BibitemShut {NoStop}%
\bibitem [{\citenamefont {{Zhu}}\ \emph {et~al.}(2016)\citenamefont {{Zhu}}
  \emph {et~al.}}]{2016MNRAS.461.1317Z}%
  \BibitemOpen
  \bibfield  {author} {\bibinfo {author} {\bibfnamefont {X.-J.}\ \bibnamefont
  {{Zhu}}} \emph {et~al.},\ }\href {\doibase 10.1093/mnras/stw1446} {\bibfield
  {journal} {\bibinfo  {journal} {\mnras}\ }\textbf {\bibinfo {volume} {461}},\
  \bibinfo {pages} {1317} (\bibinfo {year} {2016})}\BibitemShut {NoStop}%
\bibitem [{\citenamefont {{Buonanno}}(2007)}]{2007arXiv0709.4682B}%
  \BibitemOpen
  \bibfield  {author} {\bibinfo {author} {\bibfnamefont {A.}~\bibnamefont
  {{Buonanno}}},\ }\href@noop {} {\  (\bibinfo {year} {2007})},\ \Eprint
  {http://arxiv.org/abs/0709.4682} {arXiv:0709.4682 [gr-qc]} \BibitemShut
  {NoStop}%
\bibitem [{\citenamefont {Taylor}\ \emph {et~al.}(2014)\citenamefont {Taylor},
  \citenamefont {Ellis},\ and\ \citenamefont {Gair}}]{PhysRevD.90.104028}%
  \BibitemOpen
  \bibfield  {author} {\bibinfo {author} {\bibfnamefont {S.}~\bibnamefont
  {Taylor}}, \bibinfo {author} {\bibfnamefont {J.}~\bibnamefont {Ellis}}, \
  and\ \bibinfo {author} {\bibfnamefont {J.}~\bibnamefont {Gair}},\ }\href
  {\doibase 10.1103/PhysRevD.90.104028} {\bibfield  {journal} {\bibinfo
  {journal} {Phys. Rev. D}\ }\textbf {\bibinfo {volume} {90}},\ \bibinfo
  {pages} {104028} (\bibinfo {year} {2014})}\BibitemShut {NoStop}%
\bibitem [{\citenamefont {{van Haasteren}}\ \emph {et~al.}(2009)\citenamefont
  {{van Haasteren}} \emph {et~al.}}]{2009MNRAS.395.1005V}%
  \BibitemOpen
  \bibfield  {author} {\bibinfo {author} {\bibfnamefont {R.}~\bibnamefont {{van
  Haasteren}}} \emph {et~al.},\ }\href {\doibase
  10.1111/j.1365-2966.2009.14590.x} {\bibfield  {journal} {\bibinfo  {journal}
  {\mnras}\ }\textbf {\bibinfo {volume} {395}},\ \bibinfo {pages} {1005}
  (\bibinfo {year} {2009})}\BibitemShut {NoStop}%
\bibitem [{\citenamefont {Jeffreys}(1983)}]{Jeffreys}%
  \BibitemOpen
  \bibfield  {author} {\bibinfo {author} {\bibfnamefont {H.}~\bibnamefont
  {Jeffreys}},\ }\href {http://opac.inria.fr/record=b1134114} {\emph {\bibinfo
  {title} {Theory of probability}}}\ (\bibinfo  {publisher} {Clarendon Press
  New York},\ \bibinfo {address} {Oxford},\ \bibinfo {year} {1983})\BibitemShut
  {NoStop}%
\bibitem [{\citenamefont {{Feroz}}\ and\ \citenamefont
  {{Hobson}}(2008)}]{2008MNRAS.384..449F}%
  \BibitemOpen
  \bibfield  {author} {\bibinfo {author} {\bibfnamefont {F.}~\bibnamefont
  {{Feroz}}}\ and\ \bibinfo {author} {\bibfnamefont {M.~P.}\ \bibnamefont
  {{Hobson}}},\ }\href {\doibase 10.1111/j.1365-2966.2007.12353.x} {\bibfield
  {journal} {\bibinfo  {journal} {\mnras}\ }\textbf {\bibinfo {volume} {384}},\
  \bibinfo {pages} {449} (\bibinfo {year} {2008})}\BibitemShut {NoStop}%
\bibitem [{\citenamefont {Skilling}(2004)}]{JohnSkilling}%
  \BibitemOpen
  \bibfield  {author} {\bibinfo {author} {\bibfnamefont {J.}~\bibnamefont
  {Skilling}},\ }\href {\doibase 10.1063/1.1835238} {\bibfield  {journal}
  {\bibinfo  {journal} {AIP Conference Proceedings}\ }\textbf {\bibinfo
  {volume} {735}},\ \bibinfo {pages} {395} (\bibinfo {year} {2004})},\ \Eprint
  {http://arxiv.org/abs/http://aip.scitation.org/doi/pdf/10.1063/1.1835238}
  {http://aip.scitation.org/doi/pdf/10.1063/1.1835238} \BibitemShut {NoStop}%
\bibitem [{\citenamefont {{Mignard}}\ and\ \citenamefont
  {{Klioner}}(2010)}]{2010IAUS..261..306M}%
  \BibitemOpen
  \bibfield  {author} {\bibinfo {author} {\bibfnamefont {F.}~\bibnamefont
  {{Mignard}}}\ and\ \bibinfo {author} {\bibfnamefont {S.~A.}\ \bibnamefont
  {{Klioner}}},\ }\href {\doibase 10.1017/S174392130999055X} {\bibfield
  {journal} {\bibinfo  {journal} {IAU Symposium}\ }\textbf {\bibinfo {volume}
  {261}},\ \bibinfo {pages} {306} (\bibinfo {year} {2010})}\BibitemShut
  {NoStop}%
\bibitem [{\citenamefont
  {\url{https://gaia.esac.esa.int/documentation/GDR1/Catalogue_consolidation/sec_cu1cva/}}()}]{GmagHist}%
  \BibitemOpen
  \bibfield  {author} {\bibinfo {author} {\bibnamefont
  {\url{https://gaia.esac.esa.int/documentation/GDR1/Catalogue_consolidation/sec_cu1cva/}}},\
  }\href@noop {} {}\BibitemShut {NoStop}%
\bibitem [{\citenamefont {Sack}\ and\ \citenamefont
  {Urrutia}(2000)}]{Sack:2000:HCG:337150}%
  \BibitemOpen
  \bibfield  {author} {\bibinfo {author} {\bibfnamefont {J.-R.}\ \bibnamefont
  {Sack}}\ and\ \bibinfo {author} {\bibfnamefont {J.}~\bibnamefont {Urrutia}},\
  }in\ \href@noop {} {\emph {\bibinfo {booktitle} {Handbook of Computational
  Geometry}}}\ (\bibinfo  {publisher} {North-Holland Publishing Co.},\ \bibinfo
  {year} {2000})\ Chap.~\bibinfo {chapter} {5}\BibitemShut {NoStop}%
\bibitem [{\citenamefont {{Holl}}\ \emph {et~al.}(2012)\citenamefont {{Holl}}
  \emph {et~al.}}]{2012A&A...543A..15H}%
  \BibitemOpen
  \bibfield  {author} {\bibinfo {author} {\bibfnamefont {B.}~\bibnamefont
  {{Holl}}} \emph {et~al.},\ }\href {\doibase 10.1051/0004-6361/201218808}
  {\bibfield  {journal} {\bibinfo  {journal} {\aap}\ }\textbf {\bibinfo
  {volume} {543}},\ \bibinfo {eid} {A15} (\bibinfo {year} {2012})}\BibitemShut
  {NoStop}%
\end{thebibliography}%

\end{document}